\documentstyle[ltwol,epsfig,aps]{revtex}

\def\be{\begin{equation}}
\def\ee{\end{equation}}
\def\bea{\begin{eqnarray}}
\def\eea{\end{eqnarray}}

\def\ltsim{$\raisebox{0.6ex}{$<$}\!\!\!\!\!\raisebox{-0.6ex}{$\sim$}\,\,$}

\begin{document}

\title {\null\vspace*{-.0cm}\hfill  \\ \vskip
0.8cm
DISSOCIATION OF A HEAVY QUARKONIUM AT $T<T_c$}

\author{Cheuk-Yin Wong$^1$, Ted Barnes$^{1,2}$, Eric S. Swanson$^{3,4}$
and Horace W. Crater$^5$}

\address{$^1$Physics Division, Oak Ridge National Laboratory, Oak
Ridge, Tennessee 37831 USA\\ 
$^2$Department of Physics, University
of Tennessee,  Knoxville, TN 37996 USA\\
$^4$Department of Physics and Astronomy,
University of Pittsburgh, Pittsburgh, PA 15260 USA\\
$^5$Jefferson Lab, Newport News, VA 23606 USA\\
$^3$Department of Physics, University
of Tennessee Space Institute, Tullahoma, Tennessee 37388 USA
}


\twocolumn[ \maketitle\abstracts{ We examine three different ways a
heavy quarkonium can dissociate at temperatures below the
quark-gluon plasma phase transition temperature $T_c$: spontaneous
dissociation, dissociation by thermalization, and dissociation by
collision with hadrons.  We evaluate the cross sections for the
dissociation of $J/\psi$ and $\Upsilon$ in collision with $\pi$ as a
function of temperature, using the quark-interchange model of
Barnes and Swanson.  We also evaluate the dissociation temperatures
for various quarkonia, and the fraction of quarkonium lying above the
dissociation threshold as a function of temperature, using an
interquark potential inferred from lattice gauge calculations.  }]

\section{Introduction}
\vspace*{-0.1cm}

The suppression of heavy quarkonium production has been the subject of
intense interest as it was proposed as a signature for the quark-gluon
plasma \cite{Mat86}.  The recent experimental observation of an
anomalous $J/\psi$ suppression in Pb+Pb collisions by the NA50
Collaboration \cite{Gon96} has been considered by many authors
\cite{Won96}.  However, there is considerable uncertainty on the
origin of the anomalous suppression due to the lack of reliable
experimental and theoretical information on $J/\psi$ and $\chi_J$
dissociation below the quark-gluon plasma phase transition temperature
$T_c$.

We shall study three quarkonium dissociation mechanisms at $T<T_c$.  A
heavy quarkonium can dissociate spontaneously when it becomes unbound
at a temperature above its dissociation temperature.  A quarkonium
system in thermal equilibrium with the medium can dissociate by
thermalization when a fraction of the quarkonium lies at energies
above the dissociation threshold.  A heavy quarkonium can also
dissociate by collision with light hadrons. We shall discuss these
three mechanisms in turn.  Some of the details of the discussions can
be found in Refs.\ [4] and [5].

\section{Dissociation Cross Section at T=0}
\vspace*{-0.1cm}

For heavy quarkonium at $T=0$, the only dissociation mechanism for
$J/\psi$, $\chi$, and $\Upsilon$ is through their collision with
hadrons.  To date, three general approaches have been proposed for the
calculation of the dissociation cross sections: perturbative QCD
\cite{Pes79}, effective Lagrangians~\cite{Mat98}, and a
quark-interchange model using explicit quark-model wavefunctions
\cite{Won01,Mar95}.  The cross sections estimated using these approaches
vary over many orders of magnitude.

We favor the use of the quark-interchange model proposed by Barnes and
Swanson \cite{Bar92,Swa92}, as it has previously been successfully
applied to many analogous light-quark systems~\cite{Kpi} where it is
in good agreement with experiment.  The generalization of the model to
finite temperatures can also be carried out when the wave function of
the heavy quarkonium is determined as a function of
temperature~\cite{Won01c}.  A relativistic generalization of the
quark-based model has also been presented \cite{Won01b}.

To obtain the dissociation cross sections, we first use the known
hadron masses to determine interquark forces and bound state wave
functions.  We write the interaction in the form
\begin{eqnarray}
\label{eq:vpot}
V(r)
&=&{\bbox{\lambda}(i) \over 2}\cdot {\bbox{\lambda}(j) \over 2} \biggl \{
{\alpha_s \over r} - {3 b \over 4} r \nonumber\\
& &~~~ - {8 \pi \alpha_s \over
3 m_i m_j } \bbox{S}_i \cdot \bbox{S} _j \left ( {\sigma^3 \over
\pi^{3/2} } \right ) e^{-\sigma^2 r^2}  
+ V_{con}
\biggr
\}
\end{eqnarray}
where for an antiquark, the generator $\bbox{\lambda}/2$ is replaced
by $-\bbox{\lambda}^{T}/2$.  From the meson mass spectrum, we obtain the
following set of parameters:
\vspace*{-0.2cm}
\begin{eqnarray}
\label{par}
&  &\alpha_s={12 \pi \over (33-2n_f) \ln(A+Q^2/B^2)},\\
&  & A=10, ~B=0.31 {\rm ~GeV}, 
~Q^2=({\rm bound~state~mass})^2,
\nonumber\\
&  &b=0.18 {\rm ~GeV}^2, 
~\sigma=0.897 {\rm~GeV}, 
~V_{con}=0.620 {\rm ~GeV},
\nonumber\\
&  &m_u=m_d=0.334 {\rm ~ GeV}, 
~m_s=0.575 {\rm~~GeV}, 
\nonumber\\
&  &m_c=1.776 {\rm
~GeV}, 
~{\rm and}~
~m_b=5.102 {\rm~GeV}.
\nonumber
\end{eqnarray}

The dissociation of a heavy quarkonium in collision with a meson can
be described as a quark-interchange process. The Born-order scattering
amplitude can be evaluated as overlap integrals of hadron bound state
wavefunctions using the ``Feynman rules" given in Appendix C of
Ref.\ [9].

\epsfxsize=300pt
\includegraphics{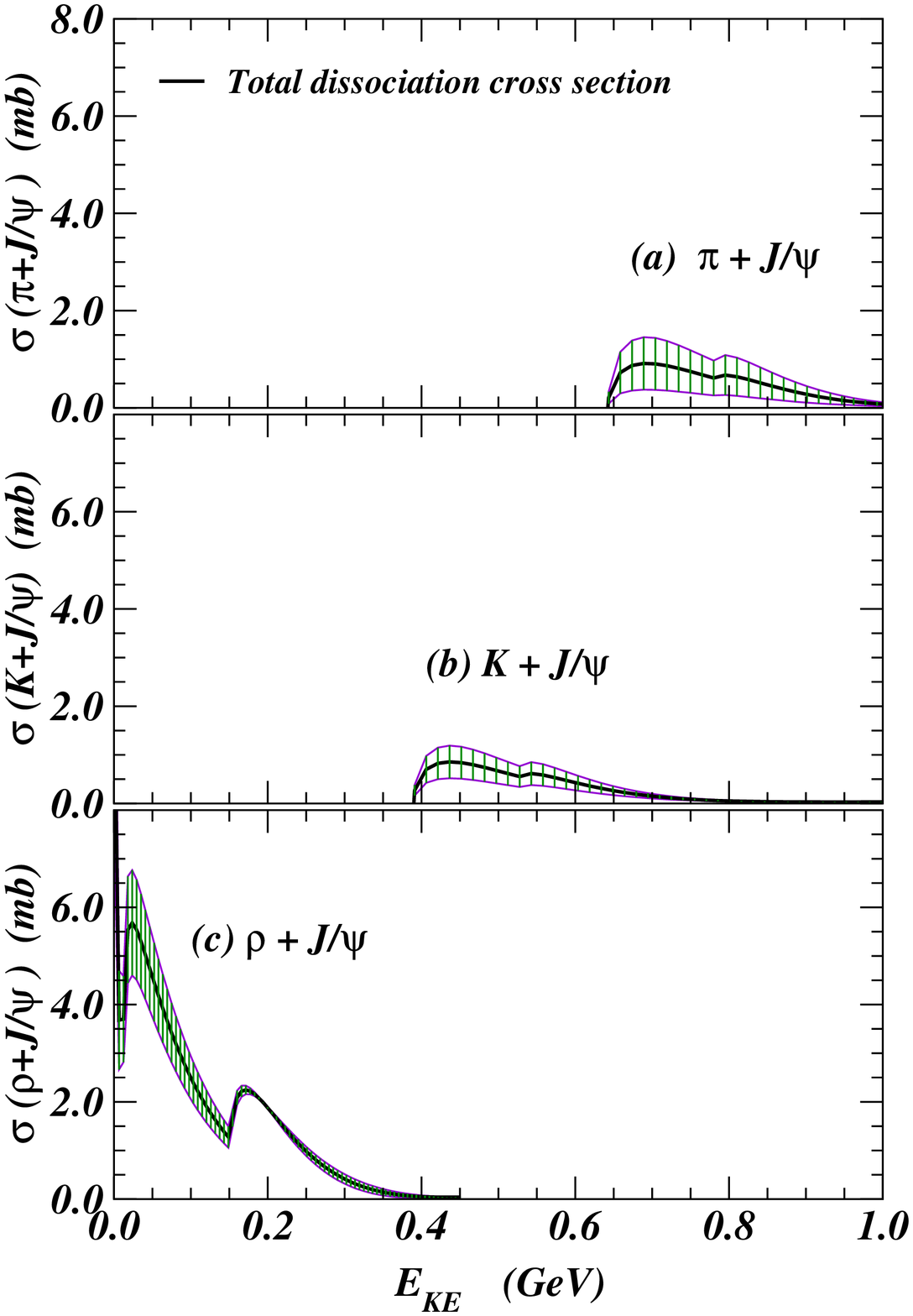}
\vspace*{+10.3cm}\hspace*{0.3cm}
\begin{minipage}[t]{7cm}
\noindent   {{\bf Fig.\ 1} Dissociation cross sections of $J/\psi$ in
collision with $\pi$, $K$, and $\rho$ at $T$ =0 
as a function of the kinetic energy $E_{KE}$. }
\end{minipage}
\vspace*{0.3cm}

We can calculate the scattering amplitude in the ``prior'' formalism
in which the interaction takes place before the interchange of the
quarks.  There is the corresponding ``post'' formalism in which the
interaction takes place after the interchange of the quarks.  In
nonrelativistic bound state scattering, one finds that these results
are actually equal, provided that the same Hamiltonian is used to
generate the asymptotic bound states and to drive the scattering
\cite{Sch68}.  We will adopt an intermediate approach and assume
relativistic kinematics and phase space but use nonrelativistic
scattering amplitudes; in consequence, we find different `post' and
`prior' cross sections in general.  Here we will take the mean value
of the `post' and `prior' results as our estimated cross section;
separate `post' and `prior' cross sections will be shown in Figs.\ 1-3
as an indication of our systematic uncertainty.

When the interaction Eq.\ (\ref{eq:vpot}) is used to calculate the
low-energy reaction, $I=2$ $\pi\pi$ scattering phase shifts,
experimental data agree well with the theoretical model.  In addition
the quark-interchange model has been applied successfully to $KN$ and
$NN$ scattering, as well as to a wide range of related no-annihilation
$K^*N$, $\Delta N$, and $\Lambda N$ reactions \cite{Kpi}.

Dissociation cross sections at $T=0$ calculated with the Barnes and
Swanson model are shown in Figs.\ 1-3.  Fig.\ 1 is for $J/\psi$, Fig.\
2 for $\chi_{c2}$, and Fig.\ 3 for $\Upsilon$.  

\epsfxsize=300pt
\includegraphics{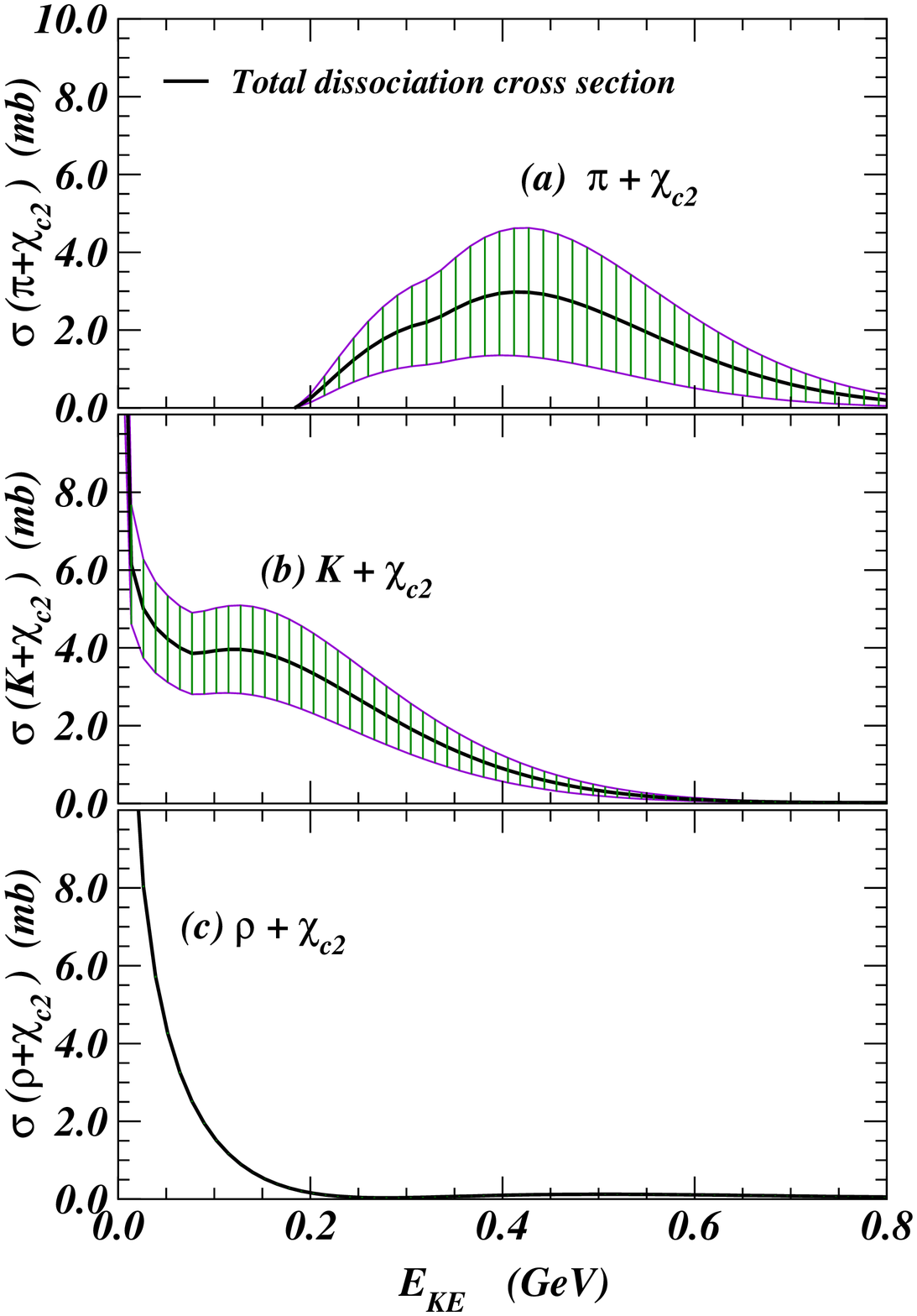}
\vspace*{9.0cm}\hspace*{0.3cm}
\begin{minipage}[t]{7cm}
\noindent   {{\bf Fig.\ 2} Dissociation cross sections of $\chi_{c2}$ in
collision with $\pi$, $K$, and $\rho$ at $T$ =0 
as a function of the kinetic energy $E_{KE}$. }
\end{minipage}
\noindent 

\vspace*{-1cm}
\epsfxsize=300pt
\includegraphics{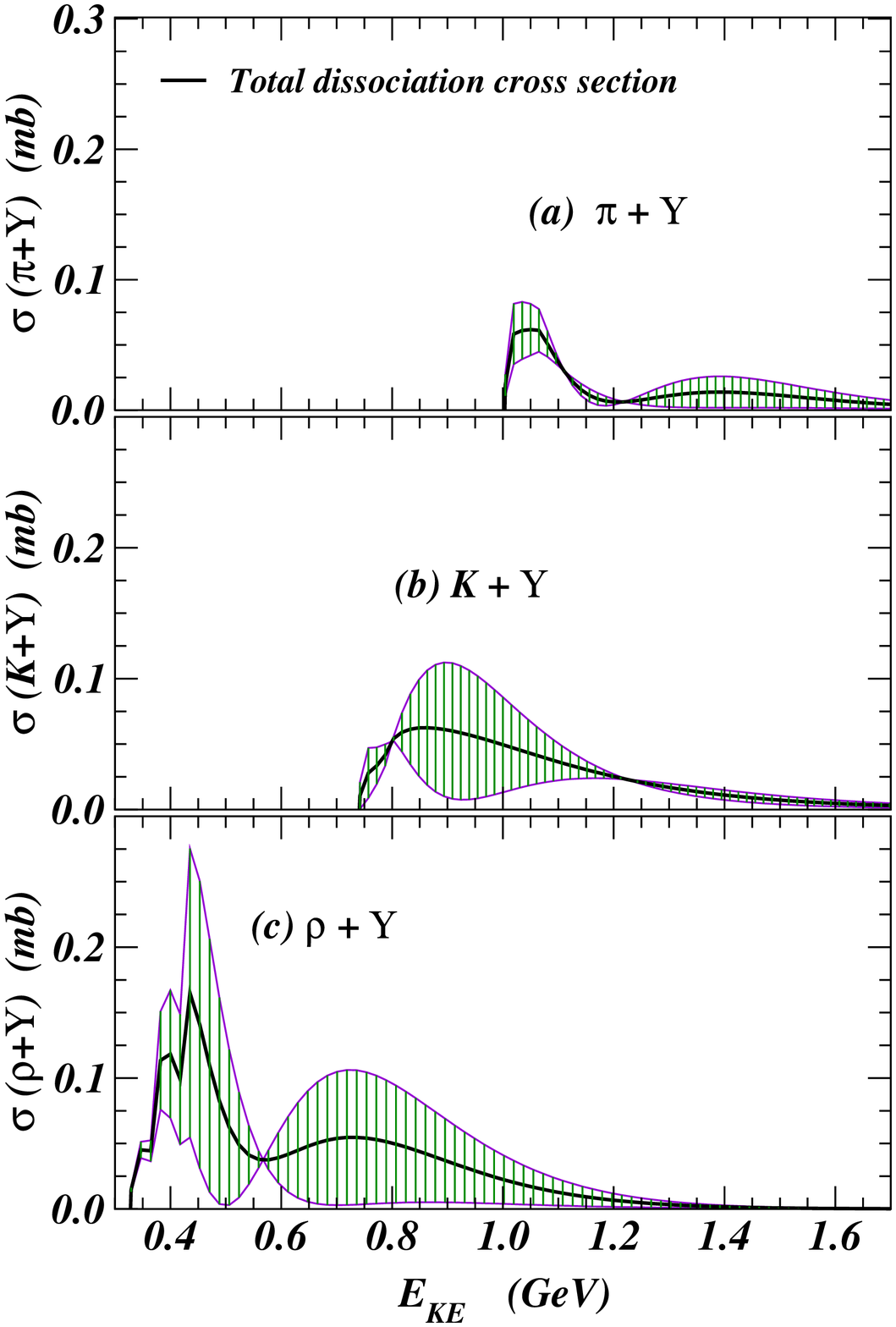}
\vspace*{+10.1cm}\hspace*{0.3cm}
\begin{minipage}[t]{7cm}
\noindent   {{\bf Fig.\ 3} Dissociation cross sections of $\Upsilon$ in
collision with $\pi$, $K$, and $\rho$ at $T$ =0 
as a function of the kinetic energy $E_{KE}$. }
\end{minipage}

One finds from Figs. 1-3 that the cross sections are sensitive
functions of energy.  The maxima of the cross sections for
$\pi+J/\psi$, $K+J/\psi$, and $\pi+\chi_{c2}$ are about 1-2 mb, and
the cross sections for exothermic reactions $\rho+J/\psi$,
$K+\chi_{c2}$, and $\rho+\chi_{c2}$ decreases from about 6 mb to about
0-2 mb as $E_{KE}$ decrease from 0.05 GeV to 0.2 GeV.  The maxima of
the cross sections of $\pi$, $K$, and $\rho$ on $\Upsilon$ are about
0.1 mb.

\section{ Heavy Quarkonium at $T<T_c$}
\vspace*{-0.1cm}

Recently, Digal, Petreczky, and Satz \cite{Dig01a} reported
theoretical results on the dissociation temperatures of heavy
quarkonia. The basic input is the temperature dependence of the
$Q$-$\bar Q$ interaction as inferred from lattice gauge calculations
of Karsch $et~al.$ \cite{Kar00}.

Due to the action of dynamical quark pairs, a proper description of
the heavy quarkonium state should be based on a screening potential
even at $T=0$ \cite{Kar88,Won99}, as the heavy quarkonium
becomes a pair of open charm or open bottom mesons when $r$ becomes
very large.

To study the behavior of a heavy quarkonium at finite temperatures, we
calculate the energy $\epsilon$ of the heavy quarkonium
single-particle state $(Q\bar Q)_{JLS}$ from the Schr\" odinger
equation \cite{Won01c}
\begin{eqnarray}
\label{eq:sch}
\biggl \{\! - \nabla \!\!\cdot\!&& { \hbar^2 \over 2\mu_{12}}\nabla   
+ V(r,T)
+(m_Q+m_{\bar Q}\nonumber\\ 
&&-M_{Q\bar q}-M_{q \bar Q})\theta (R-r)
 \!\biggr \}
\psi_{JLS} (\bbox{r})
= \epsilon \psi_{JLS} (\bbox{r}).
\end{eqnarray}
The two-body system consists of a quark pair $Q$ and $\bar Q$ at a
distance $r$\ltsim$R$ and becomes a pair of lowest-mass mesons $Q\bar
q$ and $q \bar Q$ at $r>R$.  The energy $\epsilon$ is measured
relative to the pair of lowest-mass mesons at $r\to \infty$.  The
quarkonium is bound if $\epsilon$ is negative.  It is unbound if
$\epsilon$ is positive and dissociates spontaneously, subject to
selection rules.  The quantity $\mu_{12}$ is the reduced mass.  We use
$m_Q,m_{\bar Q}, M_{Q\bar q}$, and $m_{q\bar Q}$ at $T=0$ and include
all the temperature dependence of the effective interaction in
$V(r,T)$.  For numerical calculations, $R$ has been set to 0.8 fm.

Following Karsch $et~al.$ \cite{Kar88}, we represent the $Q$-$\bar Q$
interaction by a Yukawa plus an exponential potential
\cite{Kar88,Won99}
\begin{eqnarray}
\label{eq:pot}
V(r,T)=-{4 \over 3} {\alpha_s e^{-\mu (T) r}\over r}
    -{ b(T) \over \mu(T)} e^{-\mu(T) r}
\end{eqnarray}
where $b(T)$ is the effective string-tension coefficient, $\mu(T)$ is
the effective screening parameter, and the potential is calibrated to
vanish as $r$ approaches infinity.  The results of the lattice
calculations of Karsch $et~al.$~\cite{Kar00} can be described by
$\mu(T)=\mu_0=0.28$ GeV and
\begin{eqnarray}
\label{eq:pot1}
b(T)=b_0[1-(T/T_c)^2]\theta(T_c-T),
\end{eqnarray}
where $b_0=0.35$ GeV$^2$. The value of $\mu_0$ has been fixed to be
the same as at $T=0$, and the value of $b_0$ is close to the $b$ value
obtained earlier at $T=0$ \cite{Won99}.

In our calculation, we use a running $\alpha_s$ 
as in Eq.\ (\ref{par}).  We include, in addition, the
spin-spin, spin-orbit, and tensor interactions, with a running
spin-spin interaction width as in Ref.\ [18]. 
The eigenvalues of the Hamiltonian can be obtained by matrix
diagonalization using a set of nonorthogonal Gaussian basis states
with different widths, as described in Refs.\ [10,4,5].

\section{Spontaneous Dissociation}
\vspace*{-0.1cm}

Figure 4 shows the charmonium single-particle states as a function of
temperature~\cite{Won01c}.  In considering the spontaneous
dissociation below $T_c$, it is necessary to find the selection rules
for the dissociation of a heavy quarkonium state with initial quantum
numbers $J$, $L_i$, and $S_i$ into two mesons with a total spin $S$
and a relative orbital angular momentum $L$.  With interactions of the
type in Eq.\ (\ref{eq:pot}) and the spin-spin interaction, the total
$S$ is conserved.  Parity conservation requires $\Delta L=|L-L_i|=1$
in this dissociation.  Hence, $J/\psi$ and $\psi'$ will dissociate
into a pair of open charm mesons with $L=1$, while $\chi$ states will
dissociate into a pair of open charm mesons with $L=0$ or 2.  These
conservation laws give rise to selection rules for the spontaneous
dissociation of heavy quarkonium~\cite{Won01c}.  The difference in the
energy of the lowest allowed final state and the energy of the pair of
lowest-mass meson state gives the additional increment of the
threshold energy.

\vspace*{0.5cm}
\epsfxsize=300pt
\includegraphics{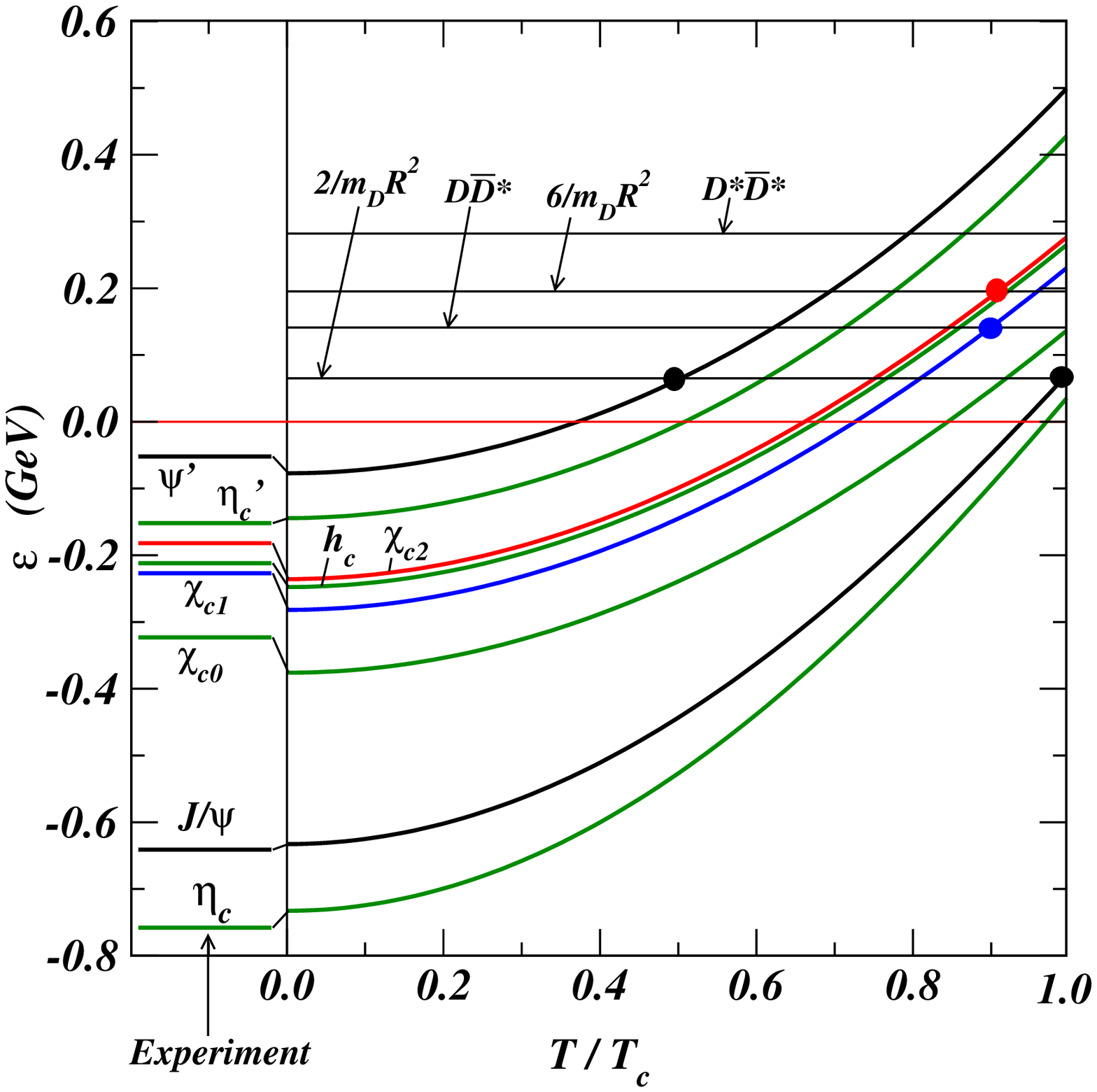}
\vspace*{+6.8cm}\hspace*{0.3cm}
\begin{minipage}[t]{7 cm}
\noindent {\bf Fig.\ 4}.  {Charmonium single-particle states as a
function of temperature.  The threshold energies are indicated as
horizontal lines.  The solid circles indicate the locations of the
dissociation temperatures. 
}
\end{minipage}
\vskip 2truemm
\noindent 

The dissociation temperatures of relevant quarkonium states can be
determined by plotting the state energies of the quarkonia and the
threshold energies after considering the selection rules.  The points
of intercept, as indicated by solid circles in Fig.\ 4, give the
positions of the dissociation temperatures.  Figure 5 shows
the bottomium single-particle states as a function of $T/T_c$.

We list the dissociation temperatures of charmonia and bottomia in
Tables I and II.  With the additional threshold energies due to the
angular momentum selection rules, the dissociation temperatures are
shifted to higher energies.  The increase in the dissociation
temperature is large for charmonia and small for bottomia.

We confirm the general features of the results of Digal $et~al.$ but
there are also differences as the dissociation temperatures depend on
the potential and the selection rules.  With the temperature-dependent
potential of Eqs.\ (\ref{eq:pot})-(\ref{eq:pot1}), the dissociation
temperatures of all heavy quarkonia, except $\chi_{b0}$, $\chi_{b1}$,
$\chi_{b2}$, and $\Upsilon$, are below $T_c$,

\epsfxsize=300pt
\includegraphics{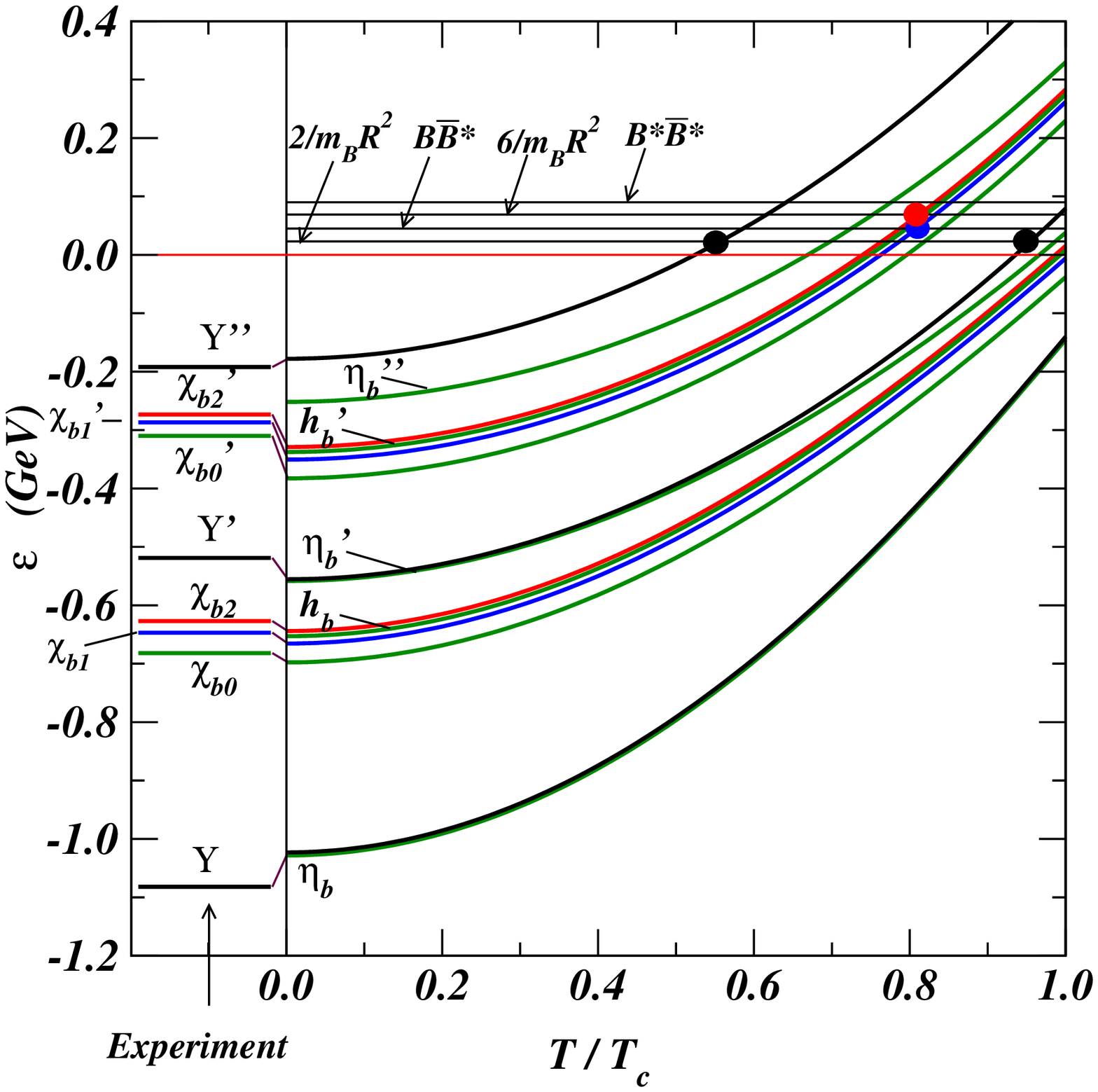}
\vspace*{+7.3cm}\hspace*{0.3cm}
\begin{minipage}[t]{7cm}
\noindent {\bf Fig.\ 5}.  {Bottomium single-particle energies as a
function of temperature.  The solid circles indicate the locations of
the dissociation temperatures. }
\end{minipage}
\vskip 4truemm
\noindent 

\begin{minipage}[t]{7.4cm}
\hspace*{0.1cm}\noindent
{\bf Table I.}  {Charmonium  dissociation temperatures $T_d$ in
units of $T_c$}
\vspace*{0.1cm}
\end{minipage}
{\vspace*{0.3cm}\hspace*{0.5cm}
\begin{tabular}{|c|c|c|c|c|}\hline
{\rm Charmonium      }& $~~\psi'~~$ &$~~\chi_{c2}~~$ 
                      & $~~\chi_{c1}~~$ & $~~J/\psi~~$ 
                        \\
\hline
 $T_d/T_c$             & 0.50 & 0.91 & 0.90&  0.99 \\
\hline  
\vspace*{-0.0cm}
$T_d/T_c$             & 0.37 & 0.66 & 0.72&  0.94 \\
\vspace*{-0.0cm}
(Selection Rules      &  &  &  & \\
not invoked)          &  &  &  & \\
\hline
\vspace*{-0.0cm}
 $T_d/T_c$         & 0.1-0.2 & 0.74 & 0.74& 1.10 \\
(Digal $et~al.$)     &  &  &  & \\
\hline
\end{tabular}
} 

\vspace*{0.4cm}
\begin{minipage}[t]{7.4cm}
\hspace*{0.1cm}\noindent
{\bf Table II.}  {Bottomium dissociation temperatures $T_d$ in
units of $T_c$}
\vspace*{0.1cm}
\end{minipage}
{\vspace*{0.3cm}\hspace*{0.0cm}
\begin{tabular}{|c|c|c|c|c|c|c|c|} \hline
{\rm Bottomium}& $\Upsilon''$ & $\chi_{b2}'$ 
                      & $\chi_{b1}'$
                      & $\Upsilon'$ & $\chi_{b2}$ 
                      & $\chi_{b1}$ & $\Upsilon$                        \\
\hline
$T_d/T_c$& 0.57 & 0.82  & 0.82  
                   & 0.96 & $>$1.0 & $>$1.0  & $>$1.0   \\
\hline  
\vspace*{-0.0cm}
$T_d/T_c$& 0.54 & 0.75  & 0.78                    
                   & 0.95 & 1.00 & $>$1.00  & $>$1.0   \\
\vspace*{-0.0cm}
(Selection Rules&  &  &
                     &  &  &  &  \\
not invoked)&  &  &
                     &  &  &  &  \\
\hline
\vspace*{-0.0cm}
$T_d/T_c$ & 0.75 & 0.83  & 0.83  
                   & 1.10 & 1.13 & 1.13  & 2.31   \\
(Digal $et~al.$)     &  &  &
                     &  &  &  &  \\
\hline
\end{tabular}
}

\vspace*{-0.1cm}
\section{Dissociation by Thermalization}
\vspace*{-0.1cm}

If the heavy quarkonium is in thermal equilibrium with the medium, the
occupation probabilities of the heavy quarkonium state $\epsilon_i$
will be distributed according to the Bose-Einstein distribution,
\begin{eqnarray}
n_i={1\over \exp\{(\epsilon_i-\mu)/T\}-1},
\end{eqnarray} 
where $\mu$ is the chemical potential.  Such a thermal equilibrium
arises from inelastic reactions of the type $h + (Q\bar Q)_{JLS} \to
h' + (Q\bar Q)_{J'L'S'}$.  When the heavy quarkonium reaches thermal
equilibrium with the medium, there is a finite fraction of the
quarkonium system to be found in excited states.  The fraction $f$ of
the system lying above their thresholds for spontaneous dissociation
will dissociate into open charm or open bottom mesons.  We call such a
process dissociation by thermalization.

\epsfxsize=300pt
\includegraphics{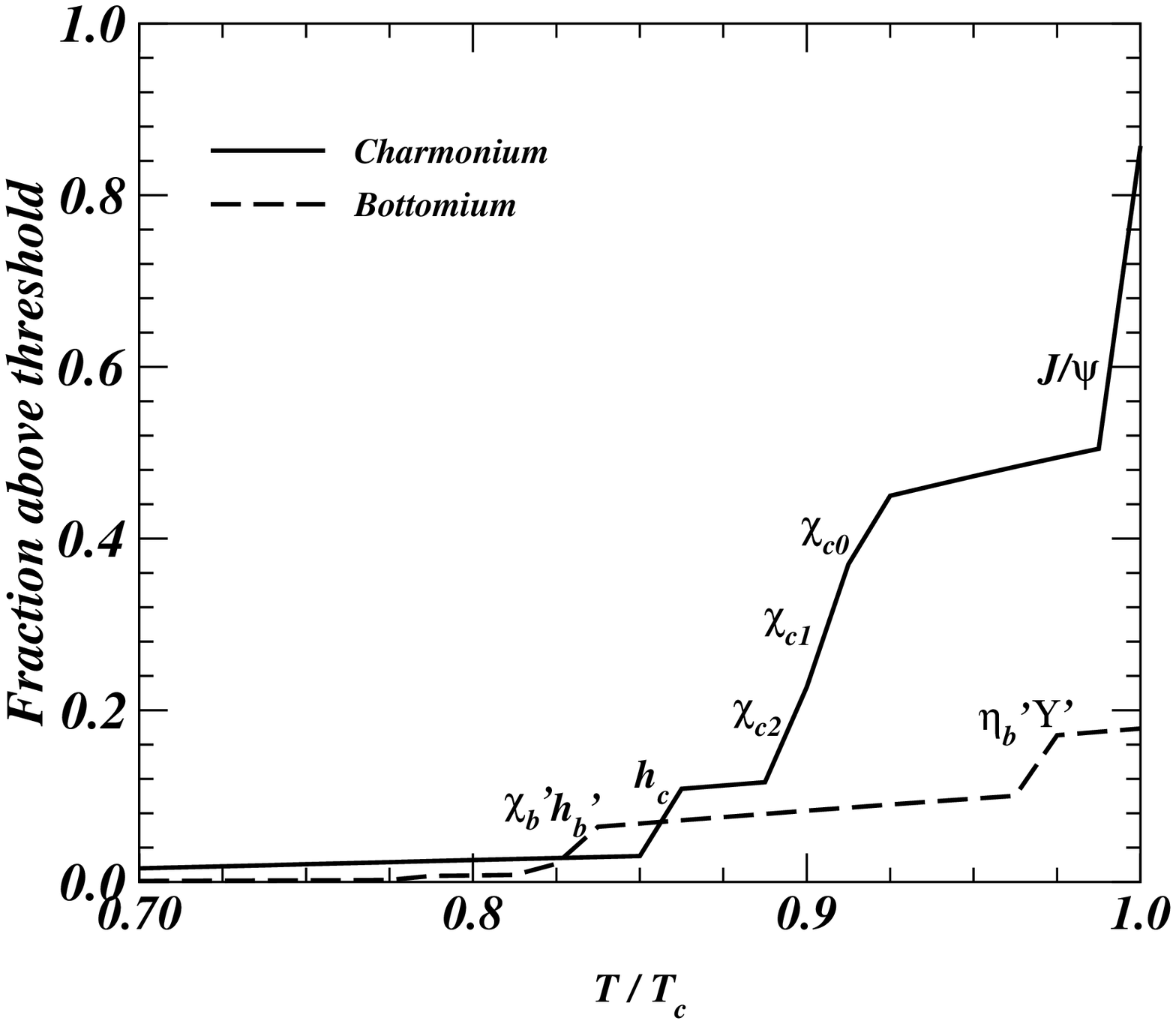}
\vspace*{+6.6cm}\hspace*{0.3cm}
\begin{minipage}[t]{7cm}
\noindent {\bf Fig.\ 6}.  {The fractions of charmonium and bottomium
lying above the dissociation threshold as a function of $T/T_c$.}
\end{minipage}
\vskip 4truemm
\noindent 

We evaluate the fraction $f$ as a function of temperature for
charmonium and bottomium, and the results are shown in Fig.\ 6.  A
state label along the curves denotes the onset of the occurrence when
that state emerges above its dissociation threshold.  As one observes,
the fraction $f$ increases with temperature.  As the temperature
approaches $T_c$, the fraction of charmonium lying above dissociation
thresholds is larger than the corresponding fraction of bottomium.

\section{Dissociation by Collision with Pions}

We study in this section the dissociation of $J/\psi$ and $\Upsilon$
by collision with pions in a medium at temperature $T$.  As the
temperature of the medium increases, the quarkonium single-particle
energy changes and the dissociation threshold energy decreases.  As a
consequence, the dissociation cross section will change. 
 
We can calculate the dissociation cross sections of $J/\psi$ and
$\Upsilon$ in collision with $\pi$ as a function of the temperature
using the Barnes and Swanson model \cite{Bar92}.  The calculation
requires the energies and wave functions of the initial and final
states, as well as the interquark interaction which leads to the
dissociation.  For the interquark interaction, we generalize the
temperature-dependent Yukawa and exponential interaction in Eqs.\
(\ref{eq:pot}) and (\ref{eq:pot1}) by replacing the color factor
$-4/3$ there with the color operator $(\bbox{\lambda}(i)/ 2)\cdot
(\bbox{\lambda}(j)/ 2)$ where for an antiquark, the generator
$\bbox{\lambda}/2$ is replaced by $-\bbox{\lambda}^{T}/2$.

Using the single-particle energies and wave functions obtained with
this temperature-dependent interaction, the scattering amplitude can
be evaluated and the dissociation cross section can be calculated, as
in Ref.\ [4].
The sum of dissociation cross sections for
$\pi+J/\psi \to D\bar D^*, D^*\bar D, D^* \bar D^* $ are shown in
Fig. 7$a$ for different temperatures $T/T_c$, as a function of the
kinetic energy $E_{KE}$. In Fig.\ 7$b$ we show similar total
dissociation cross sections for $\pi+\Upsilon \to B\bar B^*, B^*\bar
B, B^* \bar B^* $ for various temperatures.

\epsfxsize=300pt
\includegraphics{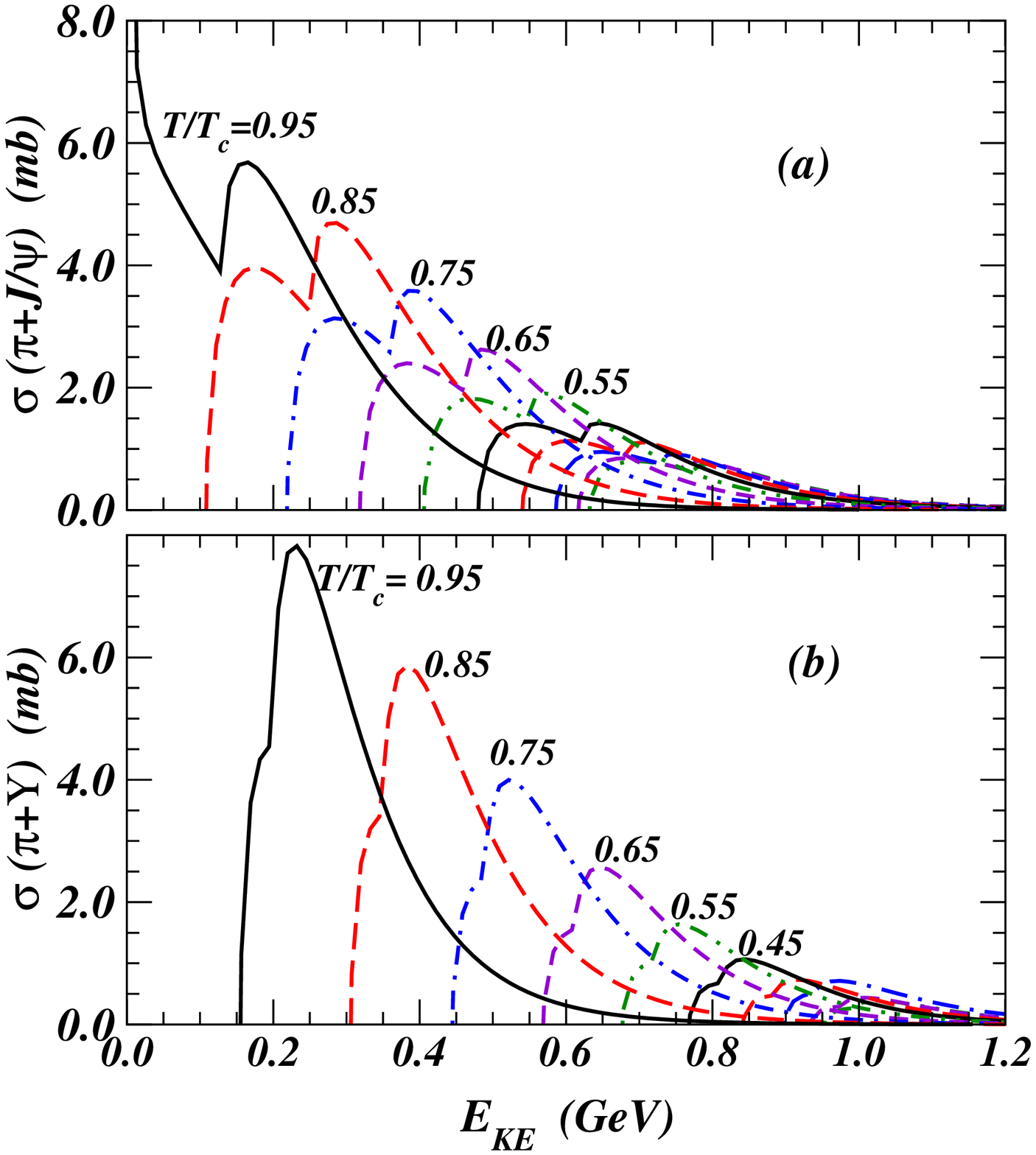}
\vspace*{+8.3cm}\hspace*{0.4cm}
\begin{minipage}[t]{7cm}
\noindent {\bf Fig.\ 7}.  {Total dissociation cross section of
$J/\psi$ (Fig.\ $a$) and $\Upsilon$ (Fig.\ $b$) in
collision with $\pi$ for various temperatures as a function of the
kinetic energy $E_{KE}$. }
\end{minipage}
\vskip 2truemm
\noindent 

We observe in Fig.\ 7 that the maximum values of the dissociation
cross sections increase and the positions of the maxima shift to lower
kinetic energies as the temperature increases. Such an increase arises
from the decrease of the threshold energies as the temperature
increases.  Over a large range of temperatures below the phase
transition temperature, dissociation cross sections of $J/\psi$ and
$\Upsilon$ in collisions with $\pi$ are large.

In a hadron gas, pions collide with a heavy quarkonium at different
energies.  We can get an idea of the energy-averaged magnitude of the
dissociation cross section by treating the pions as a Bose-Einstein
gas at temperature $T$.  In Fig. 9, we show the quantity
$\langle \sigma v \rangle $ which is the product of the dissociation
cross section and the relative velocity averaged over the energies of
the pions.  The quantity $\langle \sigma v \rangle$ is about 2 mb at
$T/T_c=0.70$ and rises to about 3 mb at $T/T_c=0.95$ where the value
of $T_c$ has been taken to be 0.175 GeV \cite{Kar00}.

\epsfxsize=300pt
\includegraphics{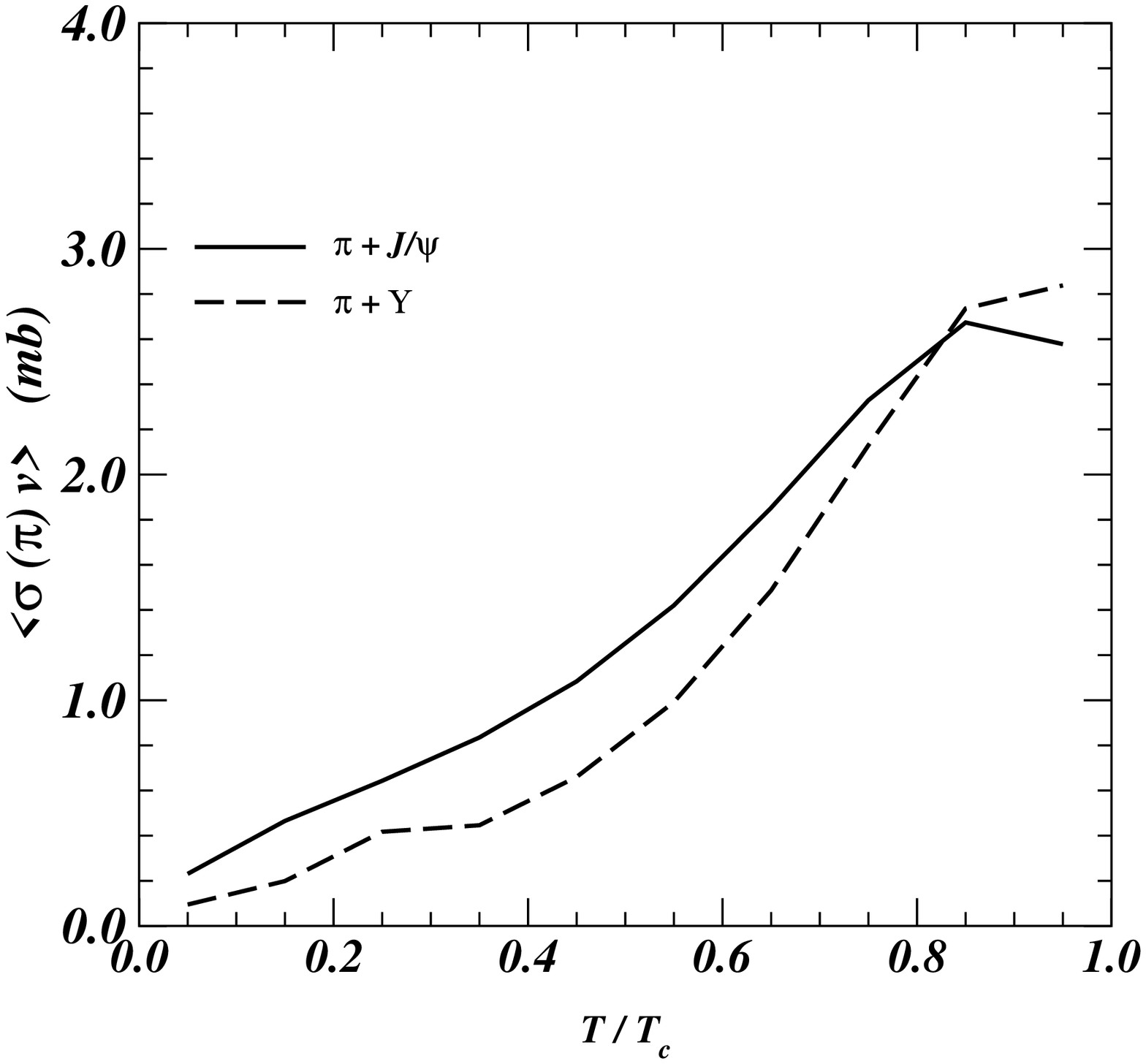}
\vspace*{+6.4cm}\hspace*{0.4cm}
\begin{minipage}[t]{7cm}
\noindent {\bf Fig.\ 8}.  {The average of the dissociation cross
section of $J/\psi$ and $\Upsilon$ in collision with $\pi$ multiplied
by $v$ as a function of the temperature.  }
\end{minipage}
\vskip 4truemm
\noindent 

We can estimate the survival probability of a heavy quarkonium in a
hot pion gas in the presence of this type of collisional dissociation.
If we represent the survival probability $S$ by $\exp\{-I\}$, the
exponential factor $I$ is given by $I= \int_{\tau_0}^{\tau_{\rm
freeze}} \langle \sigma v \rangle (\tau) \rho(\tau) d\tau $, where
$\rho(\tau)$ is the density of $\pi$ at the proper time $\tau$, and
$\tau_0$ and $\tau_{\rm freeze}$ are the initial proper time and the
freeze-out proper time respectively.  The quantity $\langle \sigma v
\rangle$ in Fig.\ 8 can be represented approximately by
\begin{eqnarray}
\langle \sigma v \rangle
= \langle \sigma v \rangle_c (T /T_c),
\end{eqnarray}
where $\langle \sigma v \rangle_c \sim 3 $ mb for $J/\psi$ or
$\Upsilon$.  Assuming a Bjorken expansion, we obtain
\begin{eqnarray}
I= \langle \sigma v \rangle_c  {dN \over dy {\cal A}}  
\left ({T_0\over T_c} \right ) 
3 \left \{ 1 - \left ( {\rho_{\rm freeze}   \tau_0 {\cal A} 
\over dN/ dy } \right ) ^{1/3}\right \},
\end{eqnarray}
where $T_0$ is the initial temperature and $\rho_{\rm freeze}$ is the
freeze-out pion density.

The degree of absorption by collision with pions depends on the
initial absorption time $\tau_0$, the freeze-out pion density
$\rho_{\rm freeze}$, and the initial temperature $T_0$.  If $T_0 \sim
T_c$, $\rho_{\rm freeze}=0.5/$ fm$^3$, and $\tau_0=3$ fm/c$+R/\gamma$,
then for the most central Pb-Pb collision at 158A GeV, the heavy
quarkonium survival probability is $S\sim 0.5$ and for the most
central Au-Au collision RHIC at $\sqrt{s_{NN}}=200$ GeV, the heavy
quarkonium survival probability is $S\sim 0.1$.  These estimates show
that the absorption of both $J/\psi$ and $\Upsilon$ by the hot pion
gas is substantial.

\section{Discussions and Conclusions}

We study the dissociation of a heavy quarkonium at $T<T_c$.  At $T =
0$, dissociation is possible only by collision with hadrons.  We
calculate the dissociation cross section for the dissociation of
$J/\psi$, $\chi$, and $\Upsilon$ in collision with $\pi$, $K$, and
$\rho$ at $T=0$.  The cross sections for $J/\psi$ and $\Upsilon$
dissociation in collision with $\pi$ are relatively small.

The temperature of the medium alters the interaction between a heavy
quark $Q$ and antiquark $\bar Q$ in the medium.  We calculate the
quarkonium single-particle states using a potential inferred from
lattice gauge calculations of Karsch $et~al.$\cite{Kar00} and obtain
the dissociation temperatures.  We confirm the general features of the
results of Digal $et~al.$ \cite{Dig01a} but there are 
differences.  We find that the selection rules change the dissociation
temperatures substantially for charmonia but only slightly for
bottomia, and that the dissociation temperatures of all heavy
quarkonia, except $\chi_{b0}$, $\chi_{b1}$, $\chi_{b2}$, and $\Upsilon$, are below
$T_c$.

A quarkonium in a medium can collide with particles in the medium to
reach thermal equilibrium.  A heavy quarkonium in thermal equilibrium
can dissociate by thermalization as there is a finite probability for
the system to be in an excited state lying above its dissociation
threshold.  We find that the fraction of the quarkonium lying above
the threshold increases with increasing temperatures.

Dissociation of a heavy quarkonium can occur in collision with
hadrons.  As the temperature increases, the threshold energies for
collisional dissociation decrease.  As a consequence, the dissociation
cross sections increase.  We have estimated the absorption of $J/\psi$
and $\Upsilon$ in collision with pions in central Pb-Pb collisions at
SPS and RHIC energies and found the absorption to be substantial.
Further microscopic investigations of the dissociation of heavy
quarkonium in collision with hadrons in high-energy heavy-ion
collisions will be of great interest.

This research was supported by the Division of Nuclear Physics,
Department of Energy, under Contract No. DE-AC05-00OR22725 managed by
UT-Battelle, LLC.  ES acknowledges support from the DOE under grant
DE-FG02-00ER41135 and DOE contract DE-AC05-84ER40150 under which the
Southeastern Universities Research Association operates the Thomas
Jefferson National Accelerator Facility.

\end{document}